\documentclass[aps,preprint]{revtex4-1}
\usepackage{graphicx}
\usepackage{epsfig}

\begin{document}

\title{Hidden Symmetries  and Geodesics of Kerr spacetime in Kaluza-Klein Theory}
\author{\large Alikram N. Aliev}
\address{Yeni Y\"{u}zy{\i}l University, Faculty of Engineering and Architecture, Cevizliba\v{g}-Topkap{\i}, 34010  Istanbul, Turkey}
\author{\large G\"{o}ksel Daylan  Esmer}
\address{Istanbul University, Department of Physics,
Vezneciler, 34134 Istanbul, Turkey}

\date{\today}

\begin{abstract}

The Kerr spacetime in Kaluza-Klein theory describes  a rotating black hole in four dimensions from  the Kaluza-Klein point of view and  involves the signature of an extra dimension that shows up through the appearance of the electric and dilaton charges. In this paper, we study  the separability  properties of the Hamilton-Jacobi equation for geodesics and the associated hidden symmetries in the spacetime of the Kerr-Kaluza-Klein black hole.
We show that the complete separation of variables occurs only for {\it massless } geodesics, implying the existence of  hidden symmetries generated by a second rank {\it conformal} Killing tensor. Employing  a simple procedure built up on an  ``effective"  metric, which is conformally related  to the original spacetime metric and admits a complete separability structure, we construct the explicit expression for the  conformal Killing tensor. Next, we study  the  properties of the geodesic motion in the equatorial plane, focusing on the cases of static and rotating Kaluza-Klein black holes separately. In both cases, we obtain the  defining equations for the  boundaries  of the regions of existence, boundedness and stability of the circular orbits as well as the analytical  formulas  for the orbital  frequency, the  radial and   vertical epicyclic frequencies of the geodesic motion.  Performing a detailed numerical analysis of these equations and frequencies, we  show that the physical effect of the extra dimension amounts to the significant  enlarging of the regions of existence, boundedness and stability towards the event horizon, regardless of the classes of orbits.

\end{abstract}

\maketitle
\newpage

\section{Introduction}

As is known, one of the most attractive features  of the ordinary Kerr spacetime, which describes a family of rotating black holes in general relativity (GR), is its separability  structure. The Hamilton-Jacobi equation for geodesics  admits  a  complete separation of variables \cite{carter1}, despite the fact that the spacetime possesses  only two global isometries generated by two commuting  Killing vector fields. Clearly,  the separability structure  implies  an extra integral of motion that in turn signals the existence of hidden symmetries in the spacetime. The authors of \cite{wp} showed that this is indeed the case. The Kerr spacetime possesses  hidden  symmetries generated by a second rank symmetric Killing tensor, rendering  the  Hamilton-Jacobi equation for geodesics completely integrable. Moreover, the Kerr spacetime  provides  full quantum separability in both the Klein-Gordon equation \cite{carter2} and  the  Dirac equation \cite{chandra, guven}. While the existence of the Killing tensor  plays a crucial role for the separability  of the Klein-Gordon equation, the situation with  the Dirac equation is more subtle. In addition to the Killing tensor, the Kerr spacetime  also admits  a second rank antisymmetric Killing-Yano tensor which can be thought of as a ``square root"  of the Killing tensor  \cite{penrose}. It is the Killing-Yano tensor that lies behind the separability of the Dirac equation  in the Kerr background \cite{cartermc}. Thus, in a sense, the usual ``square root" relationship between the Dirac and Klein-Gordon equations turns out to be echoed in the structure of hidden symmetries  of the Kerr spacetime.

In recent years, there has been an active interest in hidden symmetries of higher-dimensional black hole spacetimes. In  \cite{fk1, fk2},  it  was shown that the spacetime of higher-dimensional rotating  black holes given by  the Myers-Perry metric \cite{mp},  which  generalizes the Kerr metric to all higher dimensions, admits both the Killing and Killing-Yano tensors. In other words, the hidden symmetries  of the Kerr spacetime  survive  for the Myers-Perry spacetime in higher dimensions  as well. To gain some insight into the origin  of hidden symmetries generated by the Killing-Yano tensor, the authors of \cite{gibbons} managed to relate them to a new kind of {\it supersymmetry}, appearing in the worldline supersymmetric mechanics of spinning  point particles in the Kerr background. In a recent work \cite{ah1}, it was shown that similar analysis based on the viewpoint of worldline supersymmetric mechanics  also remains true for the  higher-dimensional spacetime of the Myers-Perry  black holes. The hidden symmetries of general rotating charged black holes  in five-dimensional minimal gauged supergravity \cite{cclp}  as well as various black hole solutions of supergravity and string theories have been studied in  a number of works (see, for instance, \cite{dkl, ad, kky, ah2, chow, finn1} and references therein).

Intriguing generalizations of the Kerr spacetime have also been studied in Kaluza-Klein theory. In a relatively simple setting, the Kerr solution in Kaluza-Klein theory describes  a rotating black hole in four dimensions from  the Kaluza-Klein point of view and  involves the signature of an extra dimension. This shows up through the appearance of the electric and dilaton charges, though the dilaton charge is not an independent parameter. That is, the solution satisfies the coupled Einstein-Maxwell-dilaton field equations, which are obtained  from the Kaluza-Klein reduction of Einstein gravity in five dimensions. The procedure of obtaining such a solution is well known
\cite{gw, fz, aht}  and amounts to boosting a four-dimensional  ``seed" solution under consideration  in the fifth dimension with a subsequent Kaluza-Klein reduction  to four dimensions. The  most general solution for rotating black  holes in Kaluza-Klein theory was obtained in \cite{rash, finn2} by  employing a solution generation technique based on the use of hidden symmetries of the Einstein field equations.

Recently, intriguing developments have also been towards exploring the  physical effects of black holes in four and higher dimensions. Observations of rapidly rotating  black holes (with the angular momenta approaching  the Kerr bound in GR) in some X-ray binaries \cite{remi, gou} have  sparked the old theoretical question of {\it bona fide spacetime geometry}  around the black holes. In light of this, many investigators have studied  gravitational effects of black holes  both in GR and beyond it, focusing  in some cases  on the imprints of the extra dimension in our physical world (for instance, see Refs.\cite{gimon, stuch1, stuch2, ap, bambi1, bambi2, ag, yazad}
and references therein).

The purpose of the present  paper is two-fold: Firstly, we examine the separability structure and the hidden symmetries  of the rotating black hole in  the Kaluza-Klein framework (the Kerr-Kaluza-Klein black hole), where it carries the imprint of the extra fifth dimension through  the electric and dilaton charges. We consider the Hamilton-Jacobi equation for  a massive (uncharged) particle in the background of this black hole and show that the complete separation  of variables occurs only for the vanishing mass of the particle, in contrast to the case of the original Kerr black hole in GR. This implies that the black hole spacetime under consideration possesses hidden symmetries generated by a second rank conformal Killing tensor. Next, we construct  the explicit  form for the  conformal Killing tensor by  employing a nice procedure  built up on an  effective metric, which is conformally related  to the original spacetime metric  and admits the separability structure due to the Killing tensor. Such a procedure of constructing the conformal Killing tensor was  earlier used in \cite{ah3} as well. Secondly,  we explore the geodesic motion of the uncharged  massive particle in the equatorial plane of the  Kerr-Kaluza-Klein black hole. Using the Hamilton-Jacobi and geodesic equations, we study the effects of the extra fifth dimension on the properties of the circular motion around this  black hole. We show that the extra dimension has its greatest effect in enlarging the regions of existence, boundedness and stability of the circular motion towards the event horizon, regardless of the classes of orbits.

The outline of the paper is as follows: In Sec.II we describe a theoretical framework for the  Kerr-Kaluza-Klein black hole. This includes a brief recalling the  construction of the pertaining spacetime metric,  the description of its physical properties as well as the properties of the Hamilton-Jacobi equation in this spacetime.  In Sec.III  we introduce a procedure that  builds up on the use of an effective metric, conformally related  to the original spacetime metric and admitting the Killing tensor. Here we show that  such a procedure enables one to construct the conformal Killing tensor for the  Kerr-Kaluza-Klein  spacetime under consideration. In Sec.IV we study the properties of the  circular and quasicircular (epicyclic) motions in the equatorial plane of both static and rotating Kaluza-Klein black holes. In both cases, we present defining equations for the  boundaries  of the existence regions  as well as for the boundaries of the regions of boundedness and stability of the circular motion. Here we also present the results of a detailed numerical analysis of these equations. Next, using the general theory of the epicyclic motion, earlier developed in \cite{ag1,ag2},  we  give the analytical expressions for the orbital, radial and vertical epicyclic frequencies and perform the numerical analysis of these expressions. In Sec.V  we conclude with the discussion of our results.

\section{The Kerr-Kaluza-Klein black hole}

We begin by recalling briefly the construction of the  exact solution that represents a  rotating  black hole in Kaluza-Klein theory, namely the Kerr-Kaluza-Klein black hole with  the Maxwell and dilaton fields. The details of the construction can be found in the original paper \cite{fz}  as well as  in a recent paper \cite{aht}, including a NUT parameter as well. At the first step, the  procedure of obtaining this solution  amounts to adding an extra spacelike flat dimension to the usual Kerr solution of four-dimensional  GR. Thus, in the Boyer-Lindquist coordinates we have the five-dimensional metric given by
 \begin{eqnarray}
ds_{5}^2 & = & -{{\Delta}\over {\Sigma}} \left(dt - a \sin^2\theta\,
d\phi \right)^2 + \Sigma \left(\frac{dr^2}{\Delta} +
d\theta^{\,2}\right)
+ \,\frac{\sin^2\theta}{\Sigma} \left[a  dt -
\left(r^2+a^2\right) d\phi\, \right]^2 + dy^2 ,
\label{5kerr}
\end{eqnarray}
where
\begin{eqnarray}
\Delta &= & r^2 - 2M r + a^2  \,, ~~~~~
\Sigma =  r^2 + a^2 \cos^2\theta \,,
\label{metfunct}
\end{eqnarray}
the parameters $ M $  and $ a $  determine the mass and the angular momentum  of the solution. Next, one needs to boost this metric in the fifth dimension by
\begin{eqnarray}
t & \rightarrow &  t \cosh\alpha + y \sinh\alpha \nonumber \\[2mm]
y & \rightarrow &  y \cosh\alpha + t \sinh\alpha\,,
\label{boost}
\end{eqnarray}
and with the velocity of the boost   $ v= \tanh\alpha $. Clearly, the boosted metric  will satisfy the vacuum equations of five-dimensional GR. Putting this metric  into the standard  Kaluza-Klein form
\begin{eqnarray}
ds_{5}^2 &=&   e^{-2 \Phi/\sqrt{3}}\,  ds_{4}^2 + e^{4 \Phi/\sqrt{3}}\,\left(dy + 2 A\right)^2,
\label{kkmetric}
\end{eqnarray}
we compactify the extra fifth dimension, identifying
the  four-dimensional metric
\begin{eqnarray}
ds_{4}^2 & = & -\frac{1}{B}\,{{\Delta}\over {\Sigma}} \left(\,dt - a \cosh\alpha \,\sin^2\theta\,d\phi\,\right)^2 + B \Sigma \left(\frac{dr^2}{\Delta}\,+\,
d\theta^{\,2}\right) -\frac{\Delta \sin^2\theta}{B} \sinh^2\alpha\, d\phi^2
\nonumber\\[2mm] &&
+ \,\frac{\sin^2\theta}{B \Sigma} \left[a dt -
\left(r^2+a^2\right)\cosh\alpha\, d\phi\, \right]^2,
 \label{solution2}
\end{eqnarray}
and  the associated potential one-form  $ A $ and  the dilaton field
$ \Phi $, which are given by
\begin{eqnarray}
A &=& \frac{Z \sinh\alpha}{2 B^2}\,\left(\cosh\alpha \, dt - a \sin^2\theta\, d\phi\right)\,,~~~~~~~~~~\Phi =  \frac{\sqrt{3}}{2}\, \ln B\,.
\label{potform1}
\end{eqnarray}
Here we have used the  notation
\begin{eqnarray}
B &=& \left(1+ \frac{2 M r \sinh^2\alpha}{\Sigma}\right)^{1/2}.
\label{b}
\end{eqnarray}
We see that for the vanishing boost velocity, $\alpha\rightarrow 0 $, the Maxwell and dilaton fields vanish and  the metric in (\ref{solution2}) reduces to the original Kerr solution. It is straightforward to check that  solution (\ref{solution2}), accompanied with  the Maxwell and dilaton fields given in (\ref{potform1}), satisfies the equation of motion derived from the four-dimensional action of Kaluza-Klein theory
\begin{eqnarray}
S &=&  \int d^4 x \sqrt{-g} \left[R - 2 \left(\partial\Phi\right)^2 - e^{2 \sqrt{3}\, \Phi} F ^2\right],
\label{4d action}
\end{eqnarray}
where  $ F=dA $. We recall that this action is obtained from the five-dimensional Einstein action for the metric in the form given by (\ref{kkmetric}). (See Ref.\cite{aht} for details).

\subsection{Physical Properties}

It is  easy to see that  the spacetime in  (\ref{solution2}) admits two commuting Killing vectors $ \,\xi_{(t)}= \partial/\partial t \, $  and $ \,\xi_{(\phi)}= \partial/\partial \phi \, $,  which reflect its time-translational and rotational invariance. Calculating the various scalar
products of these vectors,  we arrive at the  metric components in the form
\begin{eqnarray}
{\bf \xi}_{(t)} \cdot {\bf \xi}_{(t)}&=& g_{00}= - \frac{1}{B}\,\left(1 -
\frac{2 M r}{\Sigma}\right),\nonumber \\[3mm]
{\bf \xi}_{(t)} \cdot {\bf \xi}_{(\phi)}&= &g_{03}= -\, \frac{2 M r a \sin^2 \theta}{B \Sigma}\,\cosh\alpha\,,\nonumber
\\[3mm]
{\bf \xi}_{(\phi)} \cdot {\bf \xi}_{(\phi)}&=& g_{33}=
\,\left(r^2+a^2 + \frac{2 M r a^2 \sin^2 \theta}{B^2 \Sigma}
\right) B \sin^2 \theta
\,.
\label{killproduct}
\end{eqnarray}
On the other hand, as follows from  metric (\ref{solution2}), the  boosting  and dimensional reduction procedures   do not change the location of the  event horizon. It is still determined by the largest root of the equation $ \Delta =0 $, which is given by
\begin{equation}
r_{+}= M + \sqrt{M^2 - a^2 }\,,
\label{horizon1}
\end{equation}
implying that the  horizon exists provided that  $ a \leq M\, $. As for the physical parameters  of the metric, the total mass, angular momentum  and the total electric charge, they can be determined by evaluating the  corresponding Komar  integrals and the flux integral over a $2$-sphere at spatial infinity,  respectively. This has been done in works \cite{gw, fz, aht}. Writing these parameters in terms of the boost velocity $ v $, we have
\begin{eqnarray}
\mathcal{M} &=& \frac{M}{2} \left(\frac{2- v^2}{1-v^2}\right),~~~~~J = \frac{a M}{\sqrt{1-v^2}} \,\,, ~~~~~Q = \frac{M v }{1-v^2}\,.
\label{mjq1}
\end{eqnarray}
It should be noted that the dilaton charge is not independent as it can be expressed in terms of the other parameters \cite{gw, aht}. Clearly, the  ultrarelativistic limit  $ v \rightarrow 1 $  implies  the vanishing of the ``seed" (unboosted) mass $ M $ as well, thus  keeping the physical mass  $ \mathcal{M} $ fixed.

Another important feature of  spacetime (\ref{solution2})  arises from its dragging properties, which can  easily be  understood by introducing a family of  locally nonrotating observers. These observers  move on  orbits with constant $ r $ and $\theta$ and with a four-velocity $ u^{\mu} $, obeying the condition $ u\cdot{\xi}_{(\phi)}=0 $.  From this condition, we find that the  coordinate angular velocity of these observers is given by
\begin{eqnarray}
\Omega & = & -\frac{g_{03}}{g_{33}}= \frac{2 a M r \,\sqrt{1-v^2}}{2  M r(r^2+a^2) + (1-v^2) \Delta \Sigma}\,.
\label{angvelocity}
\end{eqnarray}
At large distances, we have the following expansion for the angular velocity
\begin{eqnarray}
\Omega & = & \frac{2 a M }{r^3\, \sqrt{1-v^2}} + \,\mathcal{O}\left(\frac{1}{r^4}\right)\,,
\label{drag}
\end{eqnarray}
which reveals {\it the dragging property} of metric  (\ref{solution2}) in the $\phi$-direction, vanishing at spatial infinity.  This expansion also confirms the physical angular momentum of the metric, given in (\ref{mjq1}). Meanwhile,  as follows from equation (\ref{angvelocity}), towards the event horizon  the angular velocity increases, approaching its constant value  at $  r =r_{+} $. Thus,  we have
\begin{equation}
\Omega_{H} = \frac{a}{r_{+}^2 +a^2}\, \sqrt{1-v^2}\,.
\label{horizon}
\end{equation}
It is not difficult to show that the corotating Killing  vector  defined as  $  \xi_{(t)}+ \Omega_{H}\,\xi_{(\phi)} $ is tangent to the null surface of the horizon. That is, the  quantity  $ \Omega_{H} $  is nothing but the angular velocity of the   horizon. We note that  the angular velocity  of the extreme horizon, $ a=M $, diverges in the  ultrarelativistic limit  $ v \rightarrow 1 $ as the horizon radius in  this limit  shrinks to zero, by equations (\ref{horizon1}) and (\ref{mjq1}). Therefore, in the following we will focus only on the physically acceptable  values of the boost velocity, i.e. on those obeying the condition $ v < 1 $.

In summary, the spacetime metric in (\ref{solution2}) generalizes the Kerr  solution of general relativity to include the signature of the  extra fifth dimension that in  four dimensions shows up through the appearance of the Maxwell and dilaton fields. In other words, it describes a  rotating  black hole from  the Kaluza-Klein point of view, whose  physical properties were briefly described above.

\subsection{The Hamilton-Jacobi Equation}

Let us now consider the geodesic motion of  a massive (uncharged)  particle in spacetime (\ref{solution2}) of the Kerr-Kaluza-Klein black hole. The Hamilton-Jacobi equation governing the  geodesic motion is given by
\begin{equation}
\frac{\partial S}{\partial \lambda}+\frac{1}{2}\,g^{\mu\nu}\frac{\partial S}{\partial x^\mu}\frac{\partial S}{\partial x^\nu}=0 \,,
 \label{HJeq}
\end{equation}
where $ \lambda  $ is  an affine parameter.  Since the spacetime under consideration possesses two commuting timelike and spacelike Killing vectors,
one can assume that the action  $ S $ admits the following representation
\begin{equation}
S=\frac{1}{2}\,m^2\lambda - E t+L \phi +F(r,\theta)\,.
\label{ss}
\end{equation}
Here $ F(r,\theta) $ is an arbitrary function of two variables, the constants of motion correspond to the mass  $ m $,   energy $  E $ and to the angular momentum  $ L $  of the particle.

If we now substitute this action  along with the contravariant metric components
\begin{eqnarray}
g^{00}&=&\frac{1}{B\Sigma}\left[\Sigma \sinh^2\alpha+ a^2 \sin^2\theta \cosh^2\alpha -\frac{(r^2 + a^2)^2\cosh^2\alpha}{\Delta}\right],\nonumber
\\[3mm]
g^{11}&=&\frac{\Delta}{B\Sigma}\,,~~~~~~g^{22}=\frac{1}{B\Sigma}\,,~~~~g^{03}= \frac{a \cosh\alpha}{B\Sigma}\left(1- \frac{r^2 + a^2}{\Delta}\right), \nonumber \\[3mm]
g^{33}&=& \frac{1}{B\Sigma}\left(\frac{1}{\sin^2\theta}- \frac{a^2}{\Delta}\right),
\label{contras}
\end{eqnarray}
into equation (\ref{HJeq}), it is straightforward to show that the latter  can be put in the form
\begin{eqnarray}
&& \Delta  \left(\frac{\partial F}{\partial r}\right)^2 + \left(\frac{\partial F}{\partial \theta} \right)^2 +\left[\Sigma \sinh^2 \alpha + \left(a^2\sin^2\theta - \frac{(r^2 + a^2)^2}{\Delta}\right)\cosh^2\alpha \right]E^2
\nonumber
\\[2mm] &&
+\left(\frac{1}{\sin^2\theta}- \frac{a^2}{\Delta}\right)L^2 -2\, a \cosh\alpha\left(1- \frac{r^2 + a^2}{\Delta}\right)E L = -m^2 B \Sigma\,.
\label{eq1}
\end{eqnarray}
We note that separation of $ r $  and  $\theta $ variables in this equation does not occur due to the presence of the factor $ B $ on its right-hand side (the explicit form of $ B $ is given in  Eq. (\ref{b})). On the other hand, such a separation  occurs for the particle of zero mass $(m=0)$, impliying the existence of  a new  conserved quantity, quadratic in 4-momentum, along the  null geodesics. This is due to the fact that the spacetime metric in (\ref{solution2}) admits hidden symmetries, generated by a second rank  symmetric  conformal Killing tensor \cite{wp}, which  give rise to the new  nontrivial integral of motion. Below, we  explore the hidden symmetries and present the explicit  form  for  the conformal Killing tensor.

\section{Hidden Symmetries}

We have seen that  one of the salient features of the spacetime metric in (\ref{solution2}) is that it does not allow for the complete separation of variables in the  Hamilton-Jacobi equation for massive particles. Interpreting this fact in terms of pertinent hidden symmetries, one concludes that the spacetime  does not admit hidden symmetries, which are generated by a second rank  symmetric Killing tensor $ K_{\mu\nu} $, in contrast to the original Kerr spacetime. We recall that the Killing tensor obeys the equation
\begin{equation}
\nabla_{(\lambda} K_{\mu\nu)}=0\,,
\label{Kteq}
\end{equation}
where the operator $\nabla $ denotes the covariant differentiation and the round  brackets stand for symmetrization over the indices enclosed.

Let us now assume that such a Killing tensor exists for an effective metric $ h_{\mu\nu}\, $, which is conformally related to the original  metric $ g_{\mu\nu} $ in (\ref{solution2}) as follows
\begin{eqnarray}
h_{\mu \nu} &= & e^{2\Omega} g_{\mu \nu}\,,
\label{confmet}
\end{eqnarray}
where $ \Omega $ is a smooth scalar function. It is straightforward to show that the associated Christoffel symbols $ \gamma^\lambda_{\mu \nu} $ and $ \Gamma^\lambda_{\mu \nu} $ for the metrics $ h_{\mu\nu} $ and $ g_{\mu\nu} $, respectively, are related as
\begin{eqnarray}
\gamma^\lambda_{\mu \nu}&=& \Gamma^\lambda_{\mu \nu}+ \left(\delta^\lambda_\mu \, \Omega_{,\,\nu} + \delta^\lambda_\nu \, \Omega_{,\,\mu}  - g^{\lambda\sigma}g_{\mu\nu} \, \Omega_{,\,\sigma}\right)\,.
\label{christ}
\end{eqnarray}
Meanwhile, for the covariant derivatives of a second rank symmetric tensor $ P_{\mu\nu} $  we find the relation
\begin{equation}
D_{(\lambda} P_{\mu\nu)} =\nabla_{(\lambda} P_{\mu\nu)} - 4 P_{(\mu\nu}\Omega_{,\lambda)}- g_{(\mu\nu} I_{\lambda)}\,,
\label{condeq}
\end{equation}
where
\begin{equation}
I_\lambda= -2\, g^{\alpha\tau} P_{\lambda \tau}\,\Omega_{,\alpha}\,
\label{cur}
\end{equation}
and the operator $ D $ denotes covariant differentiation with respect to the metric $ h_{\mu\nu} $,  the comma  stands for the partial derivative. It is also  straightforward  to show that for the tensor $  P_{\mu\nu} $ defined as
\begin{eqnarray}
P_{\mu \nu} &= & e^{-4\Omega} K_{\mu \nu}\,,
\label{conftens}
\end{eqnarray}
where $  K_{\mu\nu} $ is the Killing tensor in the metric   $  h_{\mu\nu} $,  equation (\ref{condeq}) reduces to the form
\begin{equation}
\nabla_{(\lambda} P_{\mu\nu)} = g_{(\mu\nu} I_{\lambda)}\,,
\label{confkil}
\end{equation}
in which we recognize the defining equation for the conformal Killing tensor $  P_{\mu\nu} $ \cite{wp}. It is worth noting that for the one-form  $ I= I_{\mu} dx^\mu  $ being exact, the conformal Killing tensor goes over into the  Killing tensor (see equation (\ref{Kteq})), implying  the representation
\begin{equation}
P_{\mu\nu} = K_{\mu\nu} + f g_{\mu\nu} \,,
\label{confkil1}
\end{equation}
where $ f $ is a scalar function.

With all this in mind, we turn now to the effective metric  $  h_{\mu\nu} $ in (\ref{confmet}). Choosing the  function  $ \Omega $  in the form
\begin{equation}
\Omega = -\, \frac{1}{2}\,\ln B\,,
\label{omega}
\end{equation}
we see that the Hamilton-Jacobi equation in the effective metric admits a complete separation of variables. In other words, in this case under consideration the factor $ B $  on the right-hand side of equation (\ref{eq1}) disappears and the resulting equation allows for separation in the  $ r $ and $\theta $ variables. Thus,  for the action $ S $ in the form
\begin{equation}
S=\frac{1}{2}m^2\lambda - E t+L \phi +S_r(r)+S_\theta(\theta)\,,
\label{ss1}
\end{equation}
we arrive at two independent ordinary differential equations
\begin{eqnarray}
\label{eqr1}
&& \Delta \left(\frac{dS_r}{d r}\right)^2 - \frac{1}{\Delta}\left[(r^2 + a^2 )\cosh\alpha \,E -a L\right]^2 + r^2 \left(m^2 +\sinh^2\alpha\, E^2 \right) = -K\,,
\\[4mm] &&
\left(\frac{dS_\theta}{d \theta}\right)^2 +\frac{1}{\sin^2\theta}\left(a \cosh\alpha\,\sin^2\theta\, E- L\right)^2 +  a^2 \cos^2\theta \left(m^2+ \sinh^2\alpha\, E^2\right)= K\,,
\label{eqth1}
\end{eqnarray}
where $ K $ is a constant of separation. It is evident that the separability occurs due to the existence of the  new quadratic integral of motion $ K=K^{\mu\nu }p_{\mu} p_{\nu} \,$,  which is guaranteed  by the existence of the irreducible Killing tensor  $  K^{\mu\nu} $  in  the effective metric $  h_{\mu\nu} $, thereby confirming our assumption made above. Using this fact in equation (\ref{eqth1})  and taking into account the relation $  m^2= -h^{\mu\nu }p_{\mu} p_{\nu} \,$, we obtain the explicit form of the Killing tensor. It is given by
\begin{eqnarray}
&&K^{\mu\nu}=  \delta^\mu_\theta \delta^\nu_\theta +  a^2 \left[ \sinh^2\alpha\,\cos^2\theta  + \cosh^2\alpha\,\sin^2\theta \right] \delta^\mu_t \delta^\nu_t
+\frac{1}{\sin^2\theta} \,\,\delta^\mu_\phi \delta^\nu_\phi  \nonumber \\[2mm] &&
+ \left(\delta^\mu_t \delta^\nu_\phi + \delta^\mu_\phi \delta^\nu_t\right) a \cosh\alpha - a^2 \cos^2\theta\, h^{\mu \nu}\,.
\label{kilcommetr}
\end{eqnarray}
For the vanishing  boost parameter,  $ \alpha \rightarrow 0 $, this expression  agrees with that obtained in  \cite{wp}  for the  ordinary Kerr spacetime.

Next, using equations (\ref{confmet}) and (\ref{conftens})  along with equation (\ref{omega}), we calculate the nonvanishing components of the ``current-vector"   in (\ref{cur}). As a consequence, we find that the associated  current one-form is given by
\begin{equation}
I = \frac{M a^2\sinh^2\alpha}{B \Sigma^2 }\left[(r^2-a^2 \cos^2\theta)\cos^2\theta\,dr + r^3 \sin 2\theta\,d\theta\right].
\label{curvect}
\end{equation}
In obtaining this expression we have also used the $ K_r^r $  and $ K_\theta^\theta $ components of the Killing tensor in (\ref{kilcommetr}). We see that this expression vanishes  for $ a=0 $  and in this case, as follows from equation (\ref{confkil}), the metric in (\ref{solution2}) admits the  reducible Killing tensor. However, in the general case it admits the conformal Killing tensor  defined in (\ref{conftens}).  Performing some calculations, we find the  explicit form  for the conformal Killing tensor
\begin{eqnarray}
P_{\mu \nu} &= & B^2  K_{\mu \nu}\,,
\label{conftens1}
\end{eqnarray}
where for the Killing tensor $ K_{\mu \nu} $  is  given by
\begin{eqnarray}
&&K_{\mu\nu} dx^{\mu}  dx^{\nu} =  - \frac{\Sigma}{\Delta}\,a^2 \cos^2\theta\, dr^2 + \frac{r^2\sin^2\theta}{B^4 \Sigma} \left[a \cosh\alpha \, dt -
(r^2+a^2)\, d\phi\, \right]^2 \nonumber \\[3mm] &&
\frac{\Delta a^2\sin^2\theta}{B^4 \Sigma}\left[\cosh\alpha \, dt -
a \sin^2\theta \, d\phi\, \right]^2 + r^2 \Sigma\, d\theta^2
+\frac{4 M r^3 \sin^2\theta \sinh^2\alpha}{B^4 \Sigma}\cdot
\nonumber \\[3mm] &&
\left[(r^2 + a^2 + M r \sinh^2\alpha) d\phi - a \,\cosh\alpha\, dt\right] d\phi\,,
\label{covkil}
\end{eqnarray}
which is obtained by lowering the contravariant indices of the tensor in (\ref{kilcommetr}) with respect to the metric $  h_{\mu\nu} $. It is straightforward to verify that the conformal Killing tensor (\ref{conftens1}) satisfies equation (\ref{confkil}) with the current-vector given in  equation (\ref{curvect}).

\section{Geodesic  Motion in the Equatorial Plane}

In this section, we restrict ourselves to the description  of  geodesics in the equatorial plane of the Kerr-Kaluza-Klein black hole. As we have mentioned above, such a black hole  transmits  the imprint of the extra fifth dimension into four-dimensional spacetime through the appearance of the electric and dilaton charges. As a consequence,  its  physical  parameters   become  substantially  different from those of the original Kerr black hole, as given in  (\ref{mjq1}).  Clearly, the effect of the extra dimension would also change  the  properties of observable orbits near the black hole.  To get some insight into this issue, it is  useful to explore the equatorial  geodesic motion  in  metric  (\ref{solution2}).  For this motion, $ \theta= \pi/2 $, from equation (\ref{eq1}) it follows  that $  \partial F/\partial \theta=0  $ and
\begin{eqnarray}
F(r)&=& \int\frac{dr}{\sqrt{\Delta}}\left\{\frac{1}{\Delta}\left[(r^2 + a^2 )\cosh\alpha \,E -a L\right]^2 - \left(a \cosh\alpha \, E- L\right)^2
\right. \nonumber \\[2mm]  & & \left.
- r^2 \left(m^2 B_0 + \sinh^2\alpha \,E\right)\right\}^{1/2},
\label{eqsr}
\end{eqnarray}
where  $ B_0 $ denotes the value of $ B $  in (\ref{b}) taken on the equatorial plane i.e.
\begin{eqnarray}
B_0 = B(r, \pi/2)= \left(1+ \frac{2 M}{r}\, \sinh^2 \alpha\right)^{1/2} . \label{b0}
\end{eqnarray}
With this in mind, using  action (\ref{ss}) in the equation
\begin{equation}
\frac{dx^\mu}{d\lambda}= g^{\mu\nu}\,\frac{\partial S}{\partial x^{\nu}}
\,\,,
\label{mom1}
\end{equation}
we obtain the following  equations of motion in the equatorial plane
\begin{eqnarray}
\Delta  B_0\,\frac{d t}{d \lambda}&=&\left[\left(r^2+a^2+\frac{2 M
a^2}{r} \right)\cosh^2 \alpha -\Delta\, \sinh^2\alpha\right]E -
\frac{2 M  a \cosh\alpha}{r}\,
L\,\,, \\[4mm]
\Delta  B_0\,
\frac{d \phi}{d \lambda}&=& \left(1-\frac{2 M
}{r}\right)L + \frac{2 M a \cosh \alpha}{r}\, E \,\,,  \\[4mm]
r^4\,B_0^2\,\left(\frac{dr}{d\lambda}\right)^2 & = & V(E,L,r,a,\alpha)\,\,,
\label{eqsmot1}
\end{eqnarray}
where the effective potential in the radial equation  (\ref{eqsmot1}) is given by
\begin{eqnarray}
V=\left[(r^2+a^2)\cosh\alpha\, E - a L\right]^2  - \Delta\,r^2 \sinh^2 \alpha\, E^2 -\Delta\left[\left(a \cosh\alpha\, E - L\right)^2 + B_0 m^2 r^2\right].\nonumber\\
\label{effpot}
\end{eqnarray}
When the right-hand side of  equation  (\ref{eqsmot1}) vanishes, the  geodesic motion occurs in circular orbits. The energy and the angular momentum of these orbits are  given by  the simultaneous solutions of the equations
\begin{eqnarray}
V&=& 0\,\,,~~~~~~~\frac{\partial V}{\partial r} =0 \,\,.
\label{simuleq1}
\end{eqnarray}
Meanwhile, the region of stability  of the circular orbits  is governed by the
inequality
\begin{equation}
\frac{\partial^2 V}{\partial r^2} \leq 0 \,\,,
\label{stability}
\end{equation}
where the case of equality refers to the innermost stable orbits.

It is worth to note  that one can also provide  an intriguing description of the equatorial motion in  black hole spacetimes by invoking the geodesic equation
\begin{equation}
\frac{d^2 x^{\mu}}{ds^2}+ \Gamma^{\mu}_{\alpha \beta} \frac{d
x^{\alpha}}{ds}\frac{d x^{\beta}}{ds}= 0 \,,
\label{eqmot}
\end{equation}
where  $ \Gamma^{\mu}_{\alpha \beta} $  are the Christoffel symbols of  the spacetime  under consideration  and  the parameter  $ s $  is supposed to be the proper time along the geodesics. Such a description  possesses some simplifying advantages, in particular, when exploring the  quasiequatorial motion by using the  method of successive approximations. In this approach, the circular motion in the equatorial plane is described  at the zeroth-order approximation. The position  four-vector of the circular orbits is given  by
\begin{eqnarray}
x_0^{\mu}(s) &= & \{t(s) \,\,,r_0\,\, ,\pi/2\,\,,\Omega_0 t(s) \},
 \label{zero1}
\end{eqnarray}
where $ \Omega_0 $ is  the  orbital frequency of the motion and it is determined by the $ \mu=1 $ component of  equation (\ref{eqmot}) on the equatorial plane. Meanwhile, the quasicircular  or  epicyclic  motion occurs due to small perturbations about  the circular orbits and  it is the subject to the first-order approximation scheme. Substituting the associated  deviation vector for small perturbations
\begin{equation}
\xi^{\mu}(s)= x^{\mu}(s)  - x_0^{\mu}(s) \,,
\label{expan}
\end{equation}
into the geodesic equation (\ref{eqmot}), we  perform an appropriate expansion in  $ \xi^{\mu} $, restricting ourselves  only to the first-order  terms. As a consequence,  we obtain the following equation
\begin{equation}
\frac{d^2 \xi^{\mu}}{dt^2} + \gamma^{\mu}_{\alpha}\,\frac{d
\xi^{\alpha}}{dt} + \xi^a \partial_a U^{\mu} = 0\,,~~~~a=1,2\equiv r,\theta \label{perteq}
\end{equation}
where the quantities   $ \gamma^{\mu}_{\alpha} $ and   $ \partial_a U^{\mu} $  are taken on a circular orbit  $r=r_0,\,\, {\theta=\pi/2}$ and  we have passed to the coordinate time $ t $, instead of the proper time $ s $. After a simple algebra, we have
\begin{eqnarray}
\gamma^{\mu}_{\alpha}&=& 2 \Gamma^{\mu}_{\alpha \beta}
u^{\beta}(u^{0})^{-1} \,,~~~~
\partial_a U^{\mu} =  \left(\partial_a \Gamma^{\mu}_{\alpha \beta}\right)
u^{\alpha} u^{\beta}
(u^{0})^{-2} \,.
\label{pertquant}
\end{eqnarray}
Next,  writing down the components of equation (\ref{perteq}), it is not difficult to  show that the epicyclic  motion consists of  two decoupled oscillations in the radial and vertical directions, which are governed by the
equations
\begin{eqnarray}
\label{radeq1}
\frac{d^2 \xi^{\,r}}{dt^2} +\Omega_{r}^2\, \xi^{\,r}&=& 0\,,~~~~~~\Omega_{r}= \left(\frac {\partial U^{r}}{\partial r}-
\gamma_{A}^r\, \gamma_{r}^A
\right)^{1/2}\,,~~~~~~~A=0, 3\equiv t,\phi\,, \\[4mm]
\frac{d^2 \xi^{\,\theta}}{dt^2} +\Omega_{\theta}^2 \,\xi^{\,\theta}
&=& 0 \,,~~~~~~~\Omega_{\theta}= \left(\frac {\partial
U^{\theta}}{\partial \theta}\right)^{1/2},
\label{verteq1}
\end{eqnarray}
respectively. It also follows that  the conditions  $\, \Omega_{r}^2 \geq 0 \,$ and $\,\Omega_{\theta}^2 \geq 0\, $ determine the stability of the circular motion against small oscillations. Thus, in this framework the description of the equatorial motion in  black hole spacetimes  can be performed in  terms of three fundamental frequencies, the orbital frequency  $\Omega_0 $, the  radial  $\Omega_r $  and  the  vertical  $\Omega_\theta $  epicyclic frequencies.

We note that the general description of the epicyclic motion in the spacetime of stationary black holes was  first given in  works \cite{ag1, ag2}. Some details of this description can also be found in recent works \cite{ap, ag}. Below,  we calculate the physical parameters of the  geodesic motion  occurring in both  equatorial and off-equatorial planes of spacetime  (\ref{solution2}),
using the  frameworks of  the Hamilton-Jacobi equation as well as the geodesic equation described above.

To make the things more transparent, it is instructive to consider the cases of static and rotating  black holes separately. In what follows,  to figure out the effects of the extra dimension, we will  express all  quantities of interest only in terms of the boost velocity $ v $ and the physical mass of the black hole. This  also  makes the things much simpler than expressing them in terms of the electric and dilaton charges.

\subsection{The static case}

Setting in equation (\ref{effpot}) the rotation parameter $ a $  to  zero
and solving the simultaneous equations in (\ref{simuleq1}), we find that the energy and the angular momentum of the circular motion around a Schwarzschild-Kaluza-Klein black hole are given by
\begin{equation}
\frac{E}{m}= \frac{1}{B_0^{1/2}}\,
\frac{(r - 2 M)\left(1 +\frac{3 M}{2r} \sinh^2\alpha\right)^{1/2}}
{\left[r^2-3 M r + M (r-4M) \sinh^2\alpha \right]^{1/2}}\,\,\,,
\label{energy}
\end{equation}
\vspace{3mm}
\begin{equation}
\frac{L}{m}= \pm\,\frac{1}{B_0^{-1/2}}\frac{M^{1/2} r^{3/2}\left(B_0^2+ \cosh^2\alpha\right)^{1/2}}{\sqrt{2} \left[r^2-3 M r + M (r-4M) \sinh^2\alpha \right]^{1/2}}\,\,.
\label{momentum}
\end{equation}
It is not difficult to see that the region of existence of the circular motion extends from infinity up to the limiting photon orbit, whose radius is governed by the vanishing  denominator of (\ref{energy}). Thus, in terms of the boost velocity $ v $ and the physical mass $ \mathcal{M} $, we have  the equation
\begin{equation}
4 \left(r- 3 \mathcal{M}\right)r - 2 \left(2 r^2 - 11 \mathcal{M} r + 8 \mathcal{M}^2\right) v^2 + \left(r- 4 \mathcal{M}\right)^2 v^4 = 0
\label{exisreg}
\end{equation}
the largest root of which is given by
\begin{equation}
\frac{r}{\mathcal{M}} = 4 + \frac{5}{v^2 - 2} + \frac{|v^2-2|}{(v^2 - 2)^2}\,\sqrt{9-8 v^2}\,.
\label{exisradi}
\end{equation}
It follows that for  $ v=0 $, the radius of the limiting  photon orbit $ r_{ph} = 3 M $ as  for the original  Schwarzschild black hole, while it moves towards the event horizon with  the growth  of $ v $ and we find that for $ v \simeq 0. 95 $, $\, r_{ph} \simeq 0.7 \mathcal{M}$. (We recall that for  $ v=0 $, we have  $\mathcal{M} = M $, as follows form  Eq.(\ref{mjq1})).

It is clear that not all  circular orbits in the region of existence  are bound. The radius of  bound circular orbits obeys  the inequality $ r > r_{mb} $,  where the radius of the innermost  bound orbit $ r_{mb} $ is given by the largest root of the equation
\begin{eqnarray}
&&\left(1 - \frac{4\mathcal{M}}{r}\, \frac{1-v^2}{2-v^2}                          \right)^2 - \left(1 + \frac{4\mathcal{M}}{r}\, \frac{v^2}{2-v^2}                          \right)^{1/2}\left[1 - \frac{4\mathcal{M}}{r}\, \frac{1-v^2}{2-v^2}
\right. \nonumber \\[3mm]  & & \left.
- \frac{\mathcal{M}}{r} \left(1 + \frac{4\mathcal{M} v^2}{r}\, \frac{1-v^2}{(2-v^2)^2}\right) \left(1 + \frac{3\mathcal{M}}{r}\, \frac{v^2}{2-v^2}\right)^{-1}
\right] = 0\,.
\label{boundradi1}
\end{eqnarray}
In obtaining this equation we have used  the condition  $ E^2=m^2$, writing the result in terms of the physical mass of the black hole. We note that for $ v=0 $, $ r_{mb} = 4 M $, while for $ v \simeq 0. 95 $, we find  that $ r_{mb} \simeq 0.8 \mathcal{M} $.

As for the region of stability of the circular motion,  its boundary is  determined by the equation
\begin{eqnarray}
&& 1 - \frac{6 \mathcal{M}}{r}\, \frac{2-3 v^2}{2- v^2} - \frac{12  \mathcal{M}^2 v^2}{r^2}\, \frac{10 v^4 - 27 v^2 +16}{(2- v^2)^3} -  \frac{16  \mathcal{M}^3 v^4}{r^3}\, \frac{1 - v^2 }{(2 - v^2)^4} \cdot
 \nonumber \\[3mm]  & &
\left(31-22 v^2 + \frac{24  \mathcal{M} v^2}{r}\, \frac{1- v^2}{2- v^2}\right)=0\,,
\label{stability1}
\end{eqnarray}
which is obtained by  using equations (\ref{energy}) and (\ref{momentum}) in  (\ref{stability}). It follows that for $ v=0 $, the radius of the innermost stable circular  orbit  $ r_{ms} = 6 M $, while for $ v \simeq  0. 95$,  we have $ r_{ms}  \simeq 1.3 \mathcal{M} $. We note that in the ultrarelativistic  limit,  $ v \rightarrow  1 $, the radius of the limiting  photon orbit as well as the radii of the innermost bound and stable orbits shrink to zero, merging with the singular horizon $  r_{+} =0 $. The results of  a detailed numerical analysis of equations (\ref{exisradi})- (\ref{stability1}) are plotted in Figure 1. We note that with increasing the boost velocity $ v $, the regions of existence, boundedness and  stability of the circular motion  essentially  enlarge towards the event horizon, thereby clearly showing up  the physical effects of the extra fifth dimension. (In this figure  and in the following ones we take $ \mathcal{M}=1 $ that makes all physical quantities of interest dimensionless).
\begin{figure}[!htmbp]%
\centering%
\includegraphics[width=10cm]{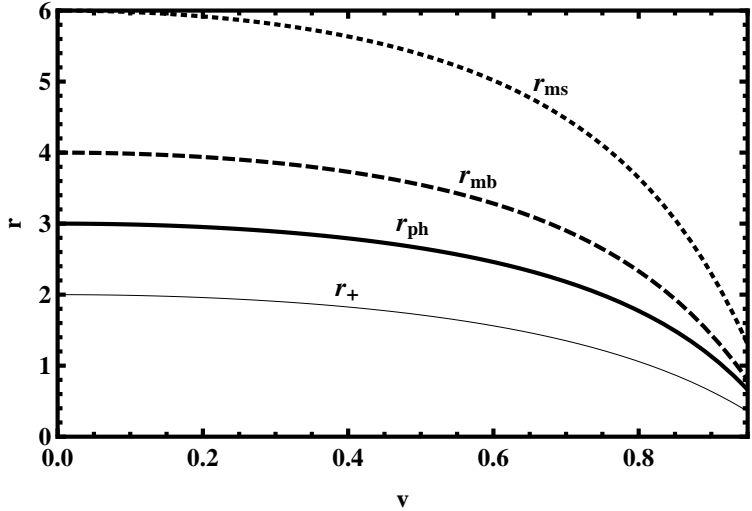}%
\caption{Radii of circular orbits around a Schwarzschid-Kaluza-Klein black hole as functions of the  boost velocity. Dotted and   dashed curves   indicate the positions of the innermost stable and  bound orbits, whereas the solid curve refers to  the limiting photon orbit. The thin curve $ r_+ $  indicates the position of the event horizon.}%
\label{Schradfreq}%
\end{figure}

Another  quantity of  physical interest  is  the binding energy  of  the innermost stable circular orbit. Using expression (\ref{energy}), it is not difficult to show that  for  $ r_{ms}  \simeq 1.3 \mathcal{M} $ and  $ v \simeq 0.95 $, the binding energy per unit  mass of a particle is
\begin{eqnarray}
E_{binding}=1-\frac{E}{m}\simeq 0.163 \,,
\end{eqnarray}
or  nearly 16.3\% of the particle rest-mass energy. That is, the energy-release process  in the vicinity of the Schwarzschild-Kaluza-Klein black hole is potentially much more efficient than for the  original Schwarzschild black hole, for which it is about  5.72\%  of  the rest-mass energy.

We turn now to the description of the equatorial motion in terms of  the orbital  and  epicyclic frequencies.  From the $ \mu=1 $ component of  equation (\ref{eqmot}) on the equatorial plane, $\theta= \pi/2 $, we find that the orbital frequency is given by
\begin{eqnarray}
\Omega_0^2 & = & \Omega_s^2 \,\left(1 + \frac{3 \mathcal{M}}{r}\frac{v^2}{2- v^2}\right)^{-1} \left(1 + \frac{4 \mathcal{M}}{r}\, \frac{v^2}{2- v^2}\right)^{-1} f(r, \mathcal{M}, v),
\label{orbitfreq1}
\end{eqnarray}
where  $ \Omega_s = \mathcal{M}^{1/2}/r^{3/2} $ is the Kepler frequency and
\begin{eqnarray}
f(r, \mathcal{M}, v) & = & 1 + \frac{4 \mathcal{M} v^2}{r}\frac{1- v^2}{(2- v^2)^2}\,.
\label{f}
\end{eqnarray}
It is not difficult to show that  using this expression for the orbital frequency in the normalization condition for the four-velocity of the particle $\, g_{\mu\nu} u^{\mu} u^{\nu}= -1 $, with equation (\ref{zero1})  and $ E= m u_0 $  in mind, we obtain the same expression  for the  energy of the circular motion as that given (\ref{energy}). Next, using equations in (\ref{radeq1}) and (\ref{verteq1})  it is straightforward to show that the vertical epicyclic frequency  $ \Omega_\theta $ is precisely the same as the orbital frequency, $ \Omega_\theta^2 =  \Omega_0^2 $, while for the radial epicyclic frequency we find that
\begin{eqnarray}
\Omega_r^2 & = & \Omega_s^2 \,\left(1 + \frac{3 \mathcal{M}}{r}\frac{v^2}{2- v^2}\right)^{-1} \left(1 + \frac{4 \mathcal{M}}{r}\, \frac{v^2}{2- v^2}\right)^{-3} h(r, \mathcal{M}, v)\,,
\label{h}
\end{eqnarray}
where the function $ h(r, \mathcal{M}, v) $ is the same as that given on the left-hand side  of equation (\ref{stability1}). It follows that the circular motion is  always stable  against  small oscillations  in the vertical direction ($\,\Omega_{\theta}^2 \geq 0\, $), while the boundary of the stability region  in the radial direction is given by the condition $ \Omega_{r}^2 = 0 $,  resulting in the same equation as in (\ref{stability1}).

\begin{figure}[!htmbp]%
\centering%
\includegraphics[width=10cm]{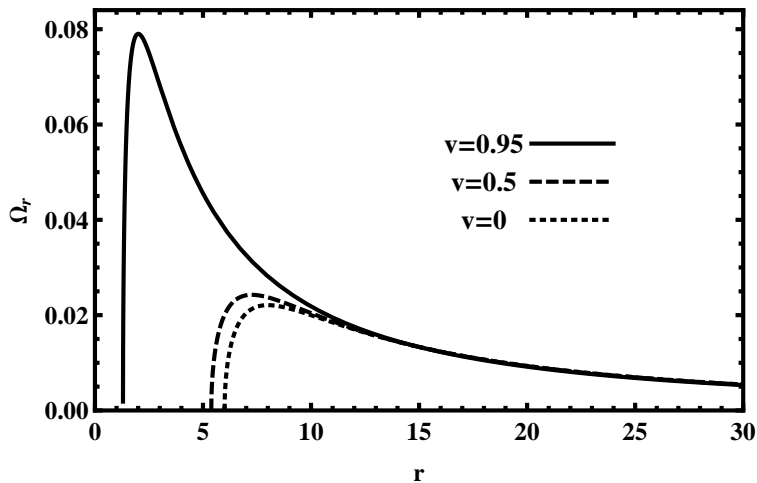}%
\caption{The dependence of the radial  epicyclic frequency on the radii of circular orbits  around a Schwarzschid-Kaluza-Klein black hole
for given values of the boost velocity.}%
\label{Schradfreq}%
\end{figure}

In Figure 2 we plot  the radial  epicyclic frequency  as a function of the radii of  circular orbits  for  given values of the boost velocity. We see that  with increasing the boost velocity, the location of the maximum  moves towards the event horizon of the black hole. Furthermore, for the large enough value of the  boost velocity, the pertaining  highest value of the radial epicyclic frequency is significantly  greater compared to that for the original Schwarzschild black hole, $ v=0 $\,.

\subsection{The rotating case}

In this case, the simultaneous solution of  the equations in (\ref{simuleq1}), determining   the energy and the angular momentum of the circular motion  turns out to be very formidable. Therefore, we appeal to the geodesic equation (\ref{eqmot}), in which case one gains some simplifying advantages. Substituting  in this equation the  Christoffel symbols  for  metric (\ref{solution2}), with  equation (\ref{zero1}) in mind,  we find that  its  $ \mu=0\,,2\,,3 \,$ components become trivial, while  the $ \mu=1 $ component yields the defining equation for the orbital frequency $ \Omega_0  $ of the circular motion. Solving this equation, we obtain that
\begin{eqnarray}
\Omega_0 & = & \frac{2  \sqrt{M} \left[- a\, \sqrt{M} (1 + X) \pm  \sqrt{X Y}\right] \sqrt{1-v^2}}{r^3 (1-v^2) X (1 + 3 X)- 2 a^2 M (1-v^2 + X)}\,,
\label{orbitfreq2}
\end{eqnarray}
where we have used the notation
\begin{eqnarray}
X & = &  1+ \frac{2 M}{r}\,\frac{v^2}{1-v^2}\,,\\
Y & = & 4 r^3 - 2 r^2 v^2 (r-5 M) + \frac{M v^4}{1-v^2} \left(3 r^2 + 6 M r -a^2\right).
\label{XY}
\end{eqnarray}
The upper sign in the numerator of (\ref{orbitfreq2}) refers to the direct orbits (the motion of the particle is corotating  with respect to the rotation of the black hole), whereas the lower sign  corresponds  to the retrograde,  counterrotating motion of the particle. Meanwhile, from the normalization condition  $ g_{\mu\nu} u^{\mu} u^{\nu}= -1, $ we find that the energy  and the orbital frequency of the circular motion are related by
\begin{equation}
\frac{E}{m}= -\frac{g_{00} + \Omega_0  g_{03}}{\sqrt{- g_{00} - 2 \Omega_0 g_{03} - \Omega_0^2 g_{33}}}\,\,,
\label{energy2}
\end{equation}
where the components of the metric tensor are given in equation (\ref{killproduct})  with  $ \theta= \pi/2 $. From this equation it follows that the radius of the limiting photon orbit is governed by the equation
\begin{equation}
1- \frac{2 M}{r} \left(1- \frac{2 a \Omega_0}{\sqrt{1-v^2}}\right) - \left(r^2+a^2 + \frac{2M}{r}\,\frac{r^2 v^2 + a^2}{1-v^2}\right) \Omega_0^2 =0\,.
\label{exist2}
\end{equation}
Next, we substitute in this equation the expression for the orbital frequency given in (\ref{orbitfreq2}) and express the result in terms of the physical mass of the black hole. Solving the resulting equation numerically, we find that in the limit of the extremal rotation, $ a \rightarrow  M $,  and for  $ v = 0 $, we have $ r_{ph} \simeq 1.23 M $  $ (a = 0.98 M) $ for the direct motion, while $  r_{ph} \simeq 4 M $   $ (a = M) $ for the retrograde motion just as for an  extreme Kerr black hole. On the other hand, for $ v = 0.95 $  and  for the the rotation parameters as given above, we find  that $ r_{ph} \simeq 0.22 \mathcal{M} $  and  $ r_{ph} \simeq 0.92 \mathcal{M} $ for  direct and retrograde orbits, respectively.

It is also not difficult to show that the radius of the innermost bound orbits  is given by the equation
\begin{eqnarray}
&& \left[1 - \frac{2 M}{r}\left(1- \frac{2 a \Omega_0}{\sqrt{1-v^2}}\right)\right] \left[1- \left(1- \frac{2 M}{r}\right) X^{-1/2}\right]
 \nonumber \\[3mm]  & &
- \Omega_0^2\left[r^2 + a^2 + \frac{2 M}{r(1-v^2)}\left(r^2 v^2 + a^2 + \frac{2 M a^2}{r} \,X^{-1/2}\right)\right]
=0\,,
\label{bound2}
\end{eqnarray}
which is obtained from equation (\ref{energy2}) with  $ E^2= m^2 $. Again, substituting  expression (\ref{orbitfreq2}) into this equation and performing the similar numerical analysis as in the case of (\ref{exist2}), we find that  for  $ v = 0 $,    $ r_{mb} \simeq 1.50 M $ (direct orbits,  $ a = 0.95 M $) and $ r_{mb} \simeq  5.83 M $ (retrograde orbits,  $ a =M $). Meanwhile, with the rotation parameters as given above and with $ v = 0.95 $,  we have $ r_{mb} \simeq 0.27 \mathcal{M} $  for direct orbits and $ r_{mb} \simeq 0.18 \mathcal{M} $  for retrograde orbits.

As in the static case, to explore the stability of the circular motion in the radial and vertical directions we need to know the explicit expressions for the pertaining epicyclic  frequencies  given in  equations (\ref{radeq1}) and (\ref{verteq1}). Using in  equation (\ref{radeq1}) the components of the Christoffel symbols  for  metric (\ref{solution2})  and performing straightforward calculations, we find that the radial epicyclic frequency is given by
\begin{eqnarray}
\Omega_r^2 & = & \frac{1}{X^3 (1-v^2)^3}\left[ \Omega_0^2 \,k_1 + \frac{2 a M \sqrt{1-v^2}}{r^3} \,\Omega_0 \, k_2 - \frac{M}{r^3}\, (1-v^2)\, k_3\right],
\label{rf2}
\end{eqnarray}
where
\begin{eqnarray}
k_1 & = &  3- \frac{8 M}{r} - \left(9 - \frac{38 M}{r} + \frac{39 M^2}{r^2}\right) v^2  + \left(1- \frac{2 M}{r}\right)^2 v^4 \left[9  - \frac{16 M}{r}
\right. \nonumber \\[2mm]  & & \left.
- \left(3- \frac{10 M}{r} + \frac{9 M^2}{r^2}\right) v^2 \right]
- \frac{M a^4}{r^5}\left[2\left(2-v^2\right)\left(1-v^2\right)^2 + \frac{3 M}{r}\, \left(3 - 4 v^2 + v^4 \right)v^2
\right. \nonumber \\[2mm]  & & \left.
+ \frac{4 M^2}{r^2}\,v^4 \right] + \frac{a^2}{r^2}\left[\left(1- \frac{10 M}{r}\right)\left(1-v^2\right)^3 + \frac{4 M^3}{r^3}\,v^2 \left(1- 8 v^2 + 5 v^4 + \frac{M}{r} \,v^2\right)
\right. \nonumber \\[2mm]  & & \left.
 + \frac{2 M^2}{r^2} \left( 1- 19  v^2 + 31 v^4 - 13  v^6 \right) \right],
\nonumber \\[3mm]
k_2 & = &  6 \left(1-v^2\right)^2 - \frac{M}{r}\,\left(2- 23 v^2 + 21 v^4\right)- \frac{4 M^2}{r^2}\,v^2\left(1-5 v^2 + \frac{M}{r}\,v^2\right)  \nonumber \\[2mm]  & &
+  \frac{a^2}{r^2} \left[4\left(1-v^2\right)^2 + \frac{9 M}{r}\,\left(1-v^2\right) v^2 + \frac{4 M^2}{r^2}\, v^4 \right],
\nonumber \\[3mm]
k_3 & = & 2 \left(1- \frac{M}{r} + \frac{2 a^2}{r^2}\right) - \left[ \frac{a^2}{r^2}\left(6- \frac{9M}{r}\right) + \left(3- \frac{8M}{r}+ \frac{4 M^2}{r^2}\right)\right]v^2
\nonumber \\[2mm]  & &
+ \left(1- \frac{2M}{r} + \frac{2 a^2}{r^2}\right)\left(1- \frac{3M}{r} + \frac{2M^2}{r^2}\right)v^4.
\label{rfcoef3}
\end{eqnarray}
For the vanishing rotation parameter, $ a  = 0 $, this expression agrees with that given in (\ref{h}) for the static black hole, whereas  for $ v = 0 $,   it  goes over into the expression for the original Kerr black hole \cite{ag1, ag2}. (See also works \cite{ap,ag}). The boundary of the stability region in the radial direction is  determined by the  equation $ \Omega_r^2 = 0 $. Writing this equation in terms of the physical mass of the black hole, we apply a numerical analysis to explore its solutions in the extremal limit of rotation $  a \rightarrow  M  $. In particular, we find that  for the direct motion and  for $ a = 0.95 M $, the radii of the  innermost  stable  orbits $ r_{ms} \simeq 1.93 M \, (v =0) $ and  $ r_{ms} \simeq 0.4  \mathcal{M} \, (v =0.95) $.  Meanwhile,  for the retrograde motion and for $ a = M $  we have $ r_{ms} = 9 M \, (v =0) $ and  $ r_{ms} \simeq 1.95  \mathcal{M} \, (v =0.95) $.

For an extreme Kerr-Kaluza-Klein black hole, the results of the full numerical analysis of the boundaries of the circular motion   are plotted in Figure 3.   The curves  clearly show  that as the boost velocity increases the radius of the limiting photon orbit as well as the radii of the innermost bound and the innermost stable orbits  essentially enlarge towards the event horizon, both  for direct  and retrograde motions.
\begin{figure}[!htmbp]%
\centering%
\includegraphics[width=10cm]{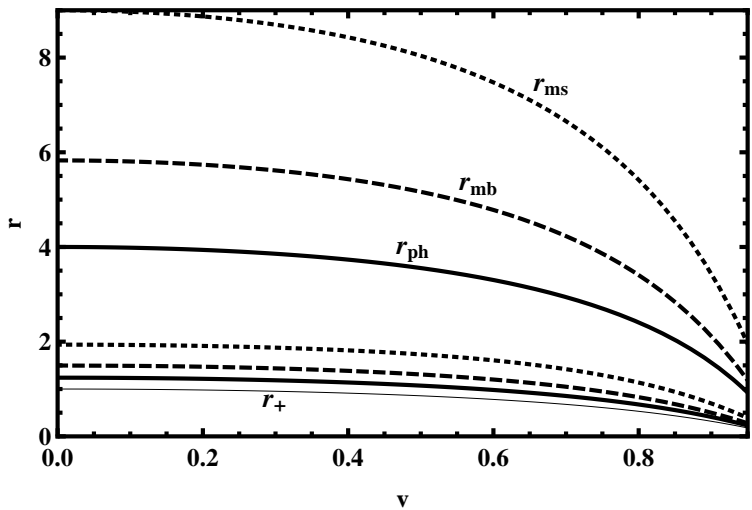}%
\caption{Radii of circular orbits around an extreme
Kerr-Kaluza-Klein black hole $(a \rightarrow  M) $ as functions of the  boost velocity. The upper set of solid, dashed and dotted curves corresponds to the limiting  photon orbit, the innermost stable and the innermost bound orbits for the retrograde  motion, respectively. Similarly,  the lower set of  solid, dashed and dotted curves refers to the limiting  photon orbit, the innermost stable and the innermost bound orbits  for the direct motion. The thin curve $ r_+ $ indicates the position of the event horizon.}%
\label{Schradfreq}%
\end{figure}

It is also of interest to explore  the dependence of the radial  epicyclic frequency on the radii of  circular orbits. In Figure 4 we illustrate  this dependence  in the limit of the extremal rotation, $ a \rightarrow  M $, and for different values of the boost velocity. We see that with increasing the boost velocity, the locations of the maxima  shift towards the event horizon for both direct and retrograde orbits. Accordingly, the pertaining values of the radial epicyclic frequency become significantly higher (especially for the retrograde  motion) compared to the case of the original Kerr black hole, $ v=0 $.
\begin{figure}[h!]
\centering
\begin{tabular}{cccc}
\epsfig{file=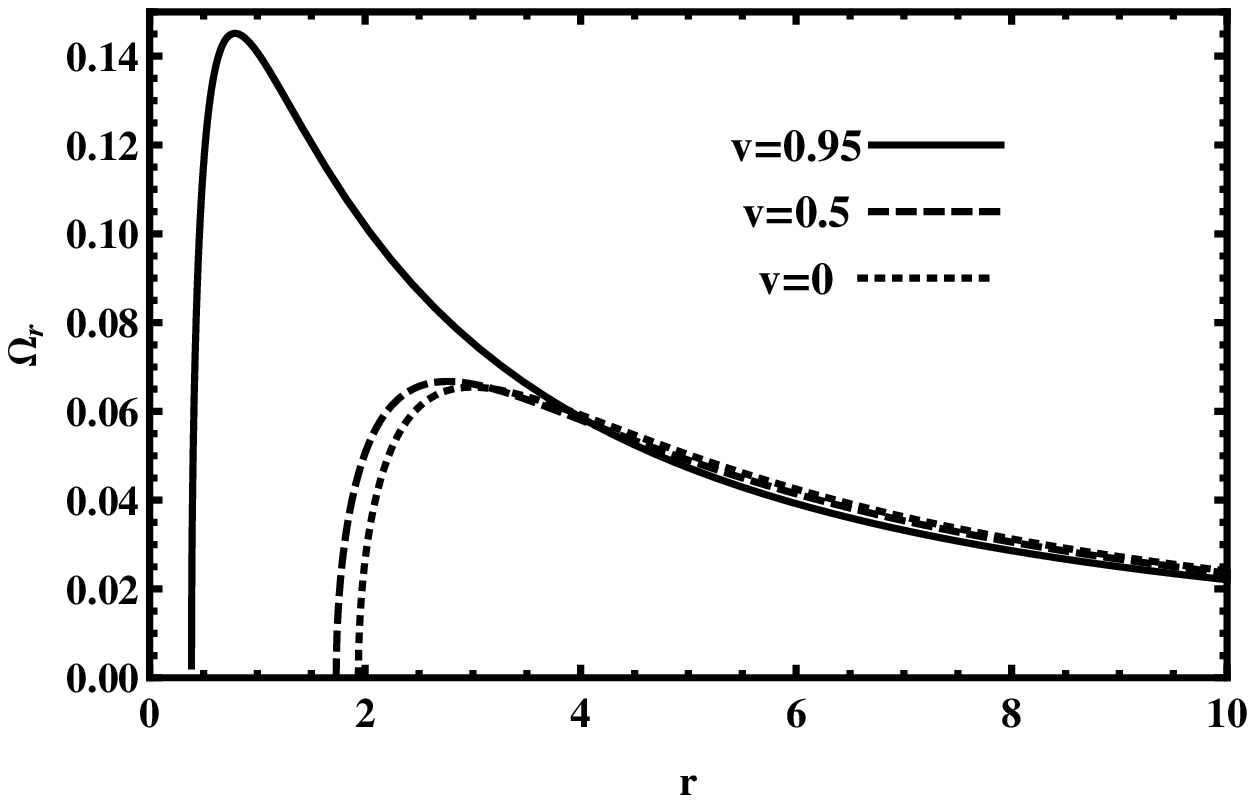,width=0.50\linewidth,clip=} &&&
\epsfig{file=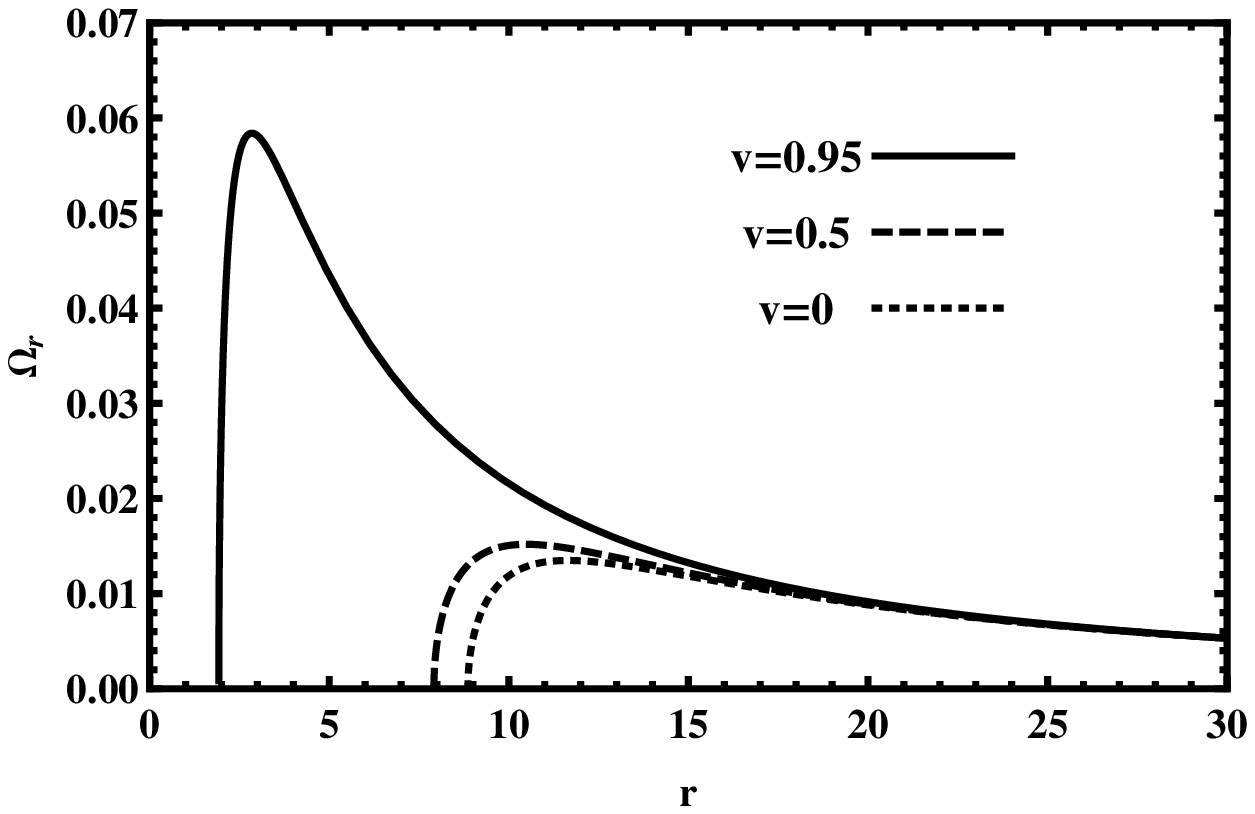,width=0.50\linewidth,clip=}
\end{tabular}
\caption{The dependence  of the radial  epicyclic frequency on the radii of circular orbits   around an extreme  Kerr-Kaluza-Klein black hole   for given values of the boost velocity. ({\it Left}\,:  Direct orbits and  $ a = 0.95 M $. {\it Right}\,: Retrograde orbits and $ a=M $.)}%
\end{figure}%

Similarly, using in equation (\ref{verteq1}) the associated Christoffel symbols  for  metric (\ref{solution2}) it is not difficult to show that the vertical epicyclic frequency is given by
\begin{eqnarray}
\Omega_{\theta}^2 & = & \frac{1}{X^2 (1-v^2)^2}\left[ \Omega_0^2 \,q_1 - \frac{4 a M \sqrt{1-v^2}}{r^3} \,\Omega_0 \, q_2 + \frac{M a^2}{r^5}\, (1-v^2)\, q_3\right],
\label{vf2}
\end{eqnarray}
where
\begin{eqnarray}
q_1 & = & X \left(1-v^2\right)\left[ X \left(1-v^2\right)+ \frac{a^2}{r^2}\left( 1-v^2 + \frac{M}{r} \,(4-v^2)\right)\right]
\nonumber \\[2mm]  & &
+ \frac{M a^4}{r^5}\left(2 - 3 v^2 + v^4 + \frac{2M}{r} \,v^2\right), \nonumber \\[3mm]
q_2 & = &  X \left(1-v^2\right) + \frac{a^2}{r^2}\left( 1-v^2 + \frac{M}{r}\,v^2\right),\nonumber
\\[2mm]
q_3 & = & 2 - v^2 + \frac{2 M}{r}\, v^2,
\label{vfcoef3}
\end{eqnarray}
It is easy to show that for $ v=0 $ this expression coincides with that for the ordinary Kerr spacetime, earlier obtained in \cite{ag1,ag2}. A detailed numerical analysis of expression (\ref{vf2}) shows that it is always nonnegative in the regions of existence and  radial stability of the circular motion. In other words, the circular  motion  is stable with respect to small perturbations in the vertical direction. In Figure 5 we plot the vertical epicyclic frequency as a function of the radii of direct orbits in the field of an extreme  Kerr-Kaluza-Klein black hole.
\begin{figure}[!htmbp]%
\centering%
\includegraphics[width=10cm]{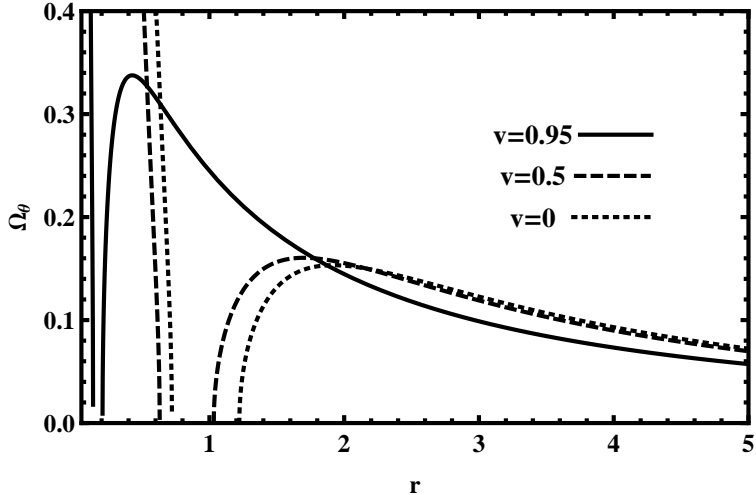}%
\caption{The dependence of the vertical  epicyclic frequency on the radii of direct circular orbits  around an extreme Kerr-Kaluza-Klein black hole $ (a = 0.95 M) $ for given values of the boost velocity.}%
\label{Schradfreq}%
\end{figure}
We note that  for the direct motion, the vertical epicyclic frequency attains its highest value in the near-horizon region. Again, the growth of the boost velocity results in  moving the location of the maxima towards the event horizon, thereby significantly increasing the maximum value of  the vertical epicyclic frequency. It is also interesting that the location of the maxima lies in the region of the radial stability of the motion. For instance, solving the  equation $\partial \Omega_{\theta}/ \partial r = 0 $ numerically,  we find that for $ a = 0.95 M $  and   $ v = 0.95  $,  the location of the maxima is given by $ r_{max} \simeq 0.42 \mathcal{M} $, which is greater than the radius of the pertaining  innermost stable circular orbit.

To conclude this subsection, we wish to calculate the binding energies of  the innermost stable circular orbits  in the field of the extreme Kerr-Kaluza-Klein black hole, for both direct and retrograde motions. Using expression (\ref{energy2}) and  performing some numerical  calculations, with equation (\ref{orbitfreq2}) in  mind, we find that  for the direct motion
\begin{eqnarray}
E_{binding}\simeq 35\,\%\,,~~~~ for~~~ a=0.95 M\,,~~~~ v=0.95\,,~~~~r_{ms}\simeq 0.4 \mathcal{M}\,,
\label{ex1}
\end{eqnarray}
in contrast to the binding energy $ E_{binding}\simeq 19\,\% $ of a particle in the Kerr field with $ a=0.95 M $ and  $ r_{ms}\simeq 1.93 \mathcal{M}\,$. Similarly, for the retrograde motion we obtain that
\begin{eqnarray}
E_{binding}\simeq 12\,\%\,,~~~~ for~~~ a= M\,,~~~~ v=0.95\,,~~~~r_{ms}\simeq 1.95 \mathcal{M}\,,
\label{ex1}
\end{eqnarray}
while in the Kerr field with  $ a=  M $ and  $ r_{ms}\simeq 9 M $, we have $ E_{binding}\simeq 3.7\,\% $. Thus, our analysis shows  that the rotating Kaluza-Klein black holes are  more energetic objects,  compared to the original Kerr black holes, in the sense of the potential energy-release process in their vicinity.

\section{Conclusion}

The remarkable property of rotating black holes in Kaluza-Klein theory is that they involve the imprint of the extra dimension  through the appearance of additional  charges in the spacetime metric. In the most simple setting, it is the Kerr spacetime that  from  the Kaluza-Klein point of view carries the signature of the extra fifth dimension by acquiring the electric and dilaton charges. In this paper, we have examined the separability  structure of the Hamilton-Jacobi equation for geodesics and the pertaining hidden symmetries in the spacetime of the Kerr-Kaluza-Klein black hole.  We have shown that in the general case of massive geodesics, the Hamilton-Jacobi equation  does not admit the complete separation of variables, whereas such a separability occurs for massless geodesics. This fact implies the existence of hidden symmetries in the spacetime, which are generated by a second rank  conformal Killing tensor.  Next, we have employed a simple framework based on the effective metric  which has the following properties: (i) it is conformally related to the original spacetime metric under consideration, (ii) it admits the Killing tensor, rendering  the associated Hamilton-Jacobi equation for massive geodesics  completely separable. With this framework, we  have constructed  the explicit expression for the conformal Killing tensor.

We have also examined the properties of the geodesic motion in the equatorial plane of the Kerr-Kaluza-Klein black holes, using  the frameworks of both the Hamilton-Jacobi and geodesic equations. In order to make the description more transparent, we have considered  the cases of static and rotating black holes separately. For both cases, we have obtained the analytical expressions for the energy and angular momentum/orbital frequency of the  circular motion  as well as we have derived the defining equations  for the  boundaries  of the regions of existence, boundedness and stability of the motion. In order to gain some simplifying advantages, we have also invoked the description of the geodesic motion in terms of three fundamental frequencies: The orbital  frequency, the  radial and   vertical epicyclic frequencies and we have obtained the associated analytical expressions for these  frequencies. Next, applying a numerical analysis,  we have found that the greatest effect of the extra fifth dimension amounts to the significant enlarging of the regions of  existence, boundedness and stability towards the event horizon, regardless of the classes of orbits.  Furthermore, it turns  out  that  for the large enough values of the  boost velocity, the locations of the maxima of the epicyclic frequencies  essentially shift towards the event   horizon, thereby resulting in much greater  values of these frequencies, compared to those for the original Schwarzschild/Kerr black holes, respectively.

Finally, we have explored the binding energy  of  the innermost stable circular orbits for both the static and rotating Kaluza-Klein  black holes. It is interesting  that for these black holes  the energy-release process in their vicinity turns out to be  potentially much more efficient than for the ordinary  Schwarzshild and Kerr black holes  of  general relativity. It should be emphasized that throughout  the paper we have  focused on the physical aspects of our description. Of course, it would also be of interest to explore  possible astrophysical implications of our results, especially in the context of high frequency quasiperiodic oscillations  observed in some black hole binaries. This is an intriguing task for future work.

\section{Acknowledgments}

One of us (A. N. Aliev)  thanks  Ekrem  \c{C}alk{\i}l{\i}\c{c} and H. H\"{u}sn\"{u} G\"{u}nd\"{u}z for their invaluable encouragement and support. He also thanks the Scientific and Technological Research Council of Turkey (T{\"U}B\.{I}TAK) for partial support under the Research Project No. 110T312. The work of G. D. E. is supported by Istanbul Univesity Scientific Research Project (BAP) No. 9227.

\end{document}